\documentclass[conference]{IEEEtran}

\usepackage{amsmath,amssymb,amsfonts}
\usepackage{textcomp}
\def\BibTeX{{\rm B\kern-.05em{\sc i\kern-.025em b}\kern-.08em
    T\kern-.1667em\lower.7ex\hbox{E}\kern-.125emX}}

\usepackage[sort, compress, numbers]{natbib}
\usepackage{xcolor}
\definecolor{navyblue}{rgb}{0.0, 0.0, 0.5}
\usepackage{hyperref}
\hypersetup{colorlinks, citecolor=blue}

\usepackage{subfigure}
\usepackage{graphicx}
\usepackage{mwe}
\usepackage{bbding}
\usepackage{subcaption}

\usepackage{algorithmicx}
\usepackage{algpseudocode}
\usepackage{algorithm}
\usepackage{tikz}
\usetikzlibrary{fit,calc}

\usepackage{tabularx}
\usepackage{tabulary}
\usepackage{booktabs}
\usepackage{multirow}
\usepackage{multicol}
\usepackage{makecell}
\usepackage{enumitem}

\usepackage{lipsum}

\newcommand{\Autoref}[1]{%
  \begingroup%
  \def\algorithmautorefname{Algorithm}%
  \def\chapterautorefname{Chapter}%
  \def\sectionautorefname{Section}%
  \def\subsectionautorefname{Section}%
  \autoref{#1}%
  \endgroup%
}

\begin{document}

\title{CUPID: A Real-Time Session-Based Reciprocal Recommendation System for a One-on-One Social Discovery Platform}

\author{
    \IEEEauthorblockN{Beomsu Kim\IEEEauthorrefmark{1}\textsuperscript{1}, 
    Sangbum Kim\IEEEauthorrefmark{1}\textsuperscript{1}, 
    Minchan Kim\IEEEauthorrefmark{1}\textsuperscript{1}, 
    Joonyoung Yi\textsuperscript{1}, 
    Sungjoo Ha\textsuperscript{1}}
    
    \IEEEauthorblockN{Suhyun Lee\textsuperscript{1}, 
    Youngsoo Lee\textsuperscript{1}, 
    Gihoon Yeom\textsuperscript{1}, 
    Buru Chang\textsuperscript{2}, 
    Gihun Lee\IEEEauthorrefmark{2}\textsuperscript{1}}
    
    \IEEEauthorblockA{\textsuperscript{1}Hyperconnect, 
    \textsuperscript{2}Sogang University}
}

\maketitle
\begingroup
    \renewcommand\thefootnote{*} 
    \footnotetext{\,These authors contributed equally to this work.}
    \renewcommand\thefootnote{\relax} 
\begingroup
    \renewcommand\thefootnote{\dag} 
    \footnotetext{\,Corresponding to: Gihun Lee (dylan.l@hpcnt.com).}
    \renewcommand\thefootnote{\relax} 
\endgroup
\endgroup

\begin{abstract}
This study introduces \textsc{Cupid}, a novel approach to session-based reciprocal recommendation systems designed for a real-time one-on-one social discovery platform. In such platforms, low latency is critical to enhance user experiences. However, conventional session-based approaches struggle with high latency due to the demands of modeling sequential user behavior for each recommendation process. Additionally, given the reciprocal nature of the platform, where users act as items for each other, training recommendation models on large-scale datasets is computationally prohibitive using conventional methods. To address these challenges, \textsc{Cupid} decouples the time-intensive user session modeling from the real-time user matching process to reduce inference time. Furthermore, \textsc{Cupid} employs a two-phase training strategy that separates the training of embedding and prediction layers, significantly reducing the computational burden by decreasing the number of sequential model inferences by several hundredfold. Extensive experiments on large-scale \textit{Azar} datasets demonstrate \textsc{Cupid}'s effectiveness in a real-world production environment. Notably, \textsc{Cupid} reduces response latency by more than 76\% compared to non-asynchronous systems, while significantly improving user engagement.
\end{abstract}


\section{Introduction}\label{sec:1_introduction}

\begin{figure*}[t]
\centering
\includegraphics[width=0.97\textwidth]{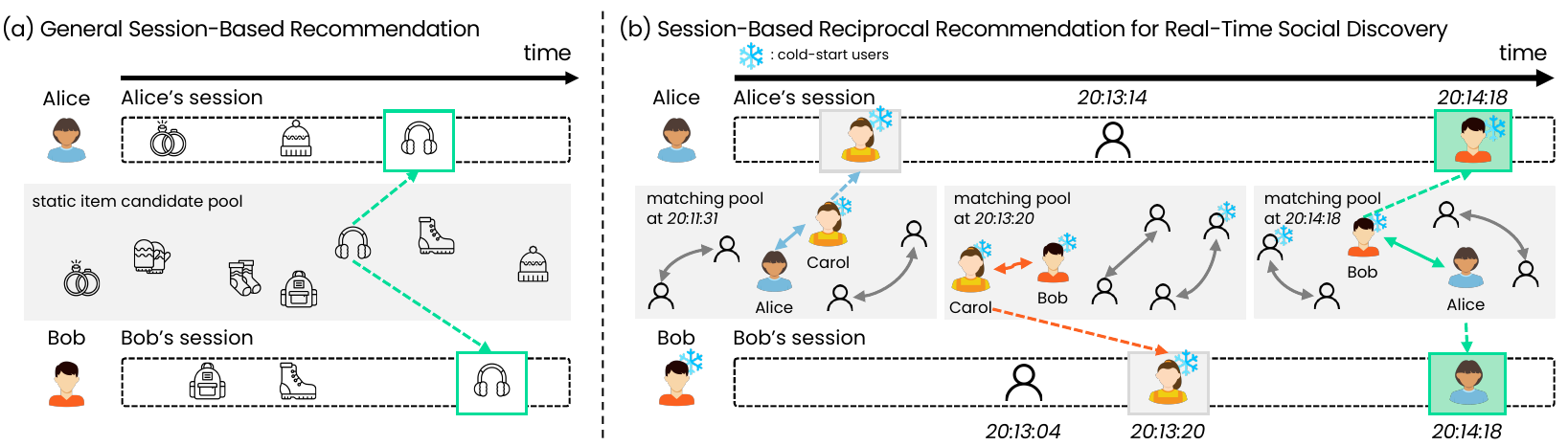}
\caption{Difference between \textbf{(a) conventional session-based recommendations} and \textbf{(b) session-based reciprocal recommendations} for real-time social discovery. The representation of the \textit{earmuffs} remains unchanged for both \textit{Alice} and \textit{Bob}. In contrast, on real-time platforms, user representations continuously evolve with each session. For instance, after \textit{Carol} interacts with both \textit{Alice} and \textit{Bob}, her representation changes based on the timing of these interactions. When \textit{Bob} pairs with \textit{Carol}, her representation now reflects her previous interaction with \textit{Alice}.}
\label{fig:1_real_time_social_discovery}
\vspace{-6pt}
\end{figure*}

\textit{Azar} is a leading real-time social discovery platform that connects users for one-on-one video conversations. To facilitate these interactions, the platform gathers users who signal their readiness for immediate video calls into a matching pool. The platform then matches users from this pool aiming to maximize overall user satisfaction, measured by the total chat duration across all pairs. Longer chat durations are indicative of more engaging and satisfying interactions, thus serving as a proxy for user satisfaction. In such reciprocal recommendation systems, where both users need to be mutually satisfied, the recommendations must reflect the preferences of both parties\,\citep{zheng2024mirror, potts2018reciprocal, zheng2018fairness}. Furthermore, as users engage with the platform, their preferences can change dynamically\,\citep{palomares2021reciprocal, pizzato2010recon}. For example, a user might start by wanting to chat casually about favorite hobbies but later seek deeper conversations about social issues.

In real-time social discovery platforms, adapting to evolving user preferences is crucial for maintaining engagement and satisfaction. One effective approach is session-based recommendations\,\citep{wang2021survey,wang2022sequential,ludewig2019performance}, where a session represents a single visit or interaction period during which the user actively engages with the platform. By focusing solely on the current session, session-based recommendations consider a user's behavior within that session rather than building a user profile from long-term historical data. This approach leverages session-specific information, enabling the system to respond to dynamic preferences\,\citep{tan2016improved,quadrana2017personalizing,liu2022monolith,hou2024dycars,mahyari2022real} and address the cold-start problem\,\citep{schein2002methods,sethi2021cold,panda2022approaches,abdullah2021eliciting,berisha2023addressing}, where new users lack sufficient historical data, by relying on data from the current session.

However, applying session-based recommendations to reciprocal recommendation systems with strict real-time constraints presents unique and significant challenges. First, conventional session-based systems build user profiles through computationally intensive session modeling\,\citep{zheng2024heterogeneous, wang2021survey, de2021transformers4rec, wang2021session}, which can take several seconds and thereby far exceeding the immediate response times required by platforms like \textit{Azar}. This delay results in a bottleneck in delivering timely recommendations. Second, user behavior in reciprocal systems can evolve rapidly within a single session, even after each interaction. For example, a positive interaction might make a user more inclined toward similar profiles, while a negative experience could shift their preferences entirely. Moreover, conventional session based recommendations mostly assume static item representations\,\citep{wang2021survey, liu2020long, hansen2020contextual}. In reciprocal systems\,\citep{cai2012reciprocal, xia2015reciprocal, potts2018reciprocal}, however, both user preferences and the items (i.e., other users) change dynamically since users act as both consumers and items. Consequently, each interaction not only updates a user's preferences but also impacts other users' representations, complicating the recommendation algorithm. These factors make real-time session-based reciprocal recommendations more complex than conventional systems. The differences between conventional session-based recommendation and its application in real-time reciprocal recommendation are illustrated in \autoref{fig:1_real_time_social_discovery}.

To address these challenges, we propose \textsc{Cupid}, a session-based reciprocal recommendation system specifically designed for real-time social discovery platforms. For the inference efficiency, \textsc{Cupid} aims to minimize the overall time consumption of the recommendation pipeline by decoupling it from the computationally intensive session modeling for each user. More specifically, \textsc{Cupid} adopts an \textit{asynchronous} session modeling approach, where user session representations are updated separately from the recommendation process. In this approach, the asynchronously updated user profiles for session modeling are stored in a separate embedding memory. On the other hand, the feature embedding, which is computationally lightweight as it relies on static user information (e.g., country, gender) or match-related statistics, is updated synchronously. When a match request arrives, the system retrieves the pre-computed session embedding from the embedding memory and combines it with the synchronously computed feature embedding to estimate the chat duration between users.

To tackle the training complexity inherent in reciprocal environments, \textsc{Cupid} divides the training process into two distinct phases. In phase 1, the focus is on training the embedding layers that model user sessions and features. In phase 2, these embedding layers are frozen, and the prediction layer is trained to estimate chat duration using the pre-trained embeddings. This two-phase strategy reduces significantly the overall computational cost, which would otherwise be much higher if both components were trained jointly. By separating the training process, the embedding layer handles each user individually rather than modeling interactions between users for every match. This approach not only lowers the training cost but also ensures high prediction performance.

In our experiments, we evaluate \textsc{Cupid} using large-scale, real-world data from \textit{Azar}. Both offline and online production tests demonstrate that \textsc{Cupid} significantly reduces the recommendation latency and enhances overall user satisfaction, proving its effectiveness for real-time reciprocal recommendation systems. Notably, implementing \textsc{Cupid} increases the average chat duration by 6.8\% for warm-start users and 5.9\% for cold-start users. At the same time, it reduces response latency by 79.7\% for the 90-th percentile of users and 75.9\% for the 99-th percentile in the \textit{Azar} service.

\vspace{3pt}
\noindent Our main contributions are summarized as follows:

\begin{itemize}
    \vspace{1.5pt}
    \item We systematically formulate session-based reciprocal recommendation systems for real-time social discovery platforms. To the best of our knowledge, this is the first study to tackle this specific challenge. \textbf{(\Autoref{sec:problem_formulation})}

    \vspace{4pt}
    \item We introduce \textsc{Cupid}, a novel session-based recommendation system for real-time reciprocal recommendation. Using asynchronous session embedding and a two-phase training strategy, \textsc{Cupid} improves both inference time and training efficiency. \textbf{(\Autoref{sec:cupid})}

    \vspace{4pt}
    \item We validate the efficacy of \textsc{Cupid} using large-scale real-world data from \textit{Azar}. \textsc{Cupid} significantly enhances recommendation performance in both offline and online evaluations while meeting strict latency constraints required in real-time social discovery. \textbf{(\Autoref{sec:experiments})}
\end{itemize}
\begin{figure*}[t]
\centering
\includegraphics[width=0.85\textwidth]
{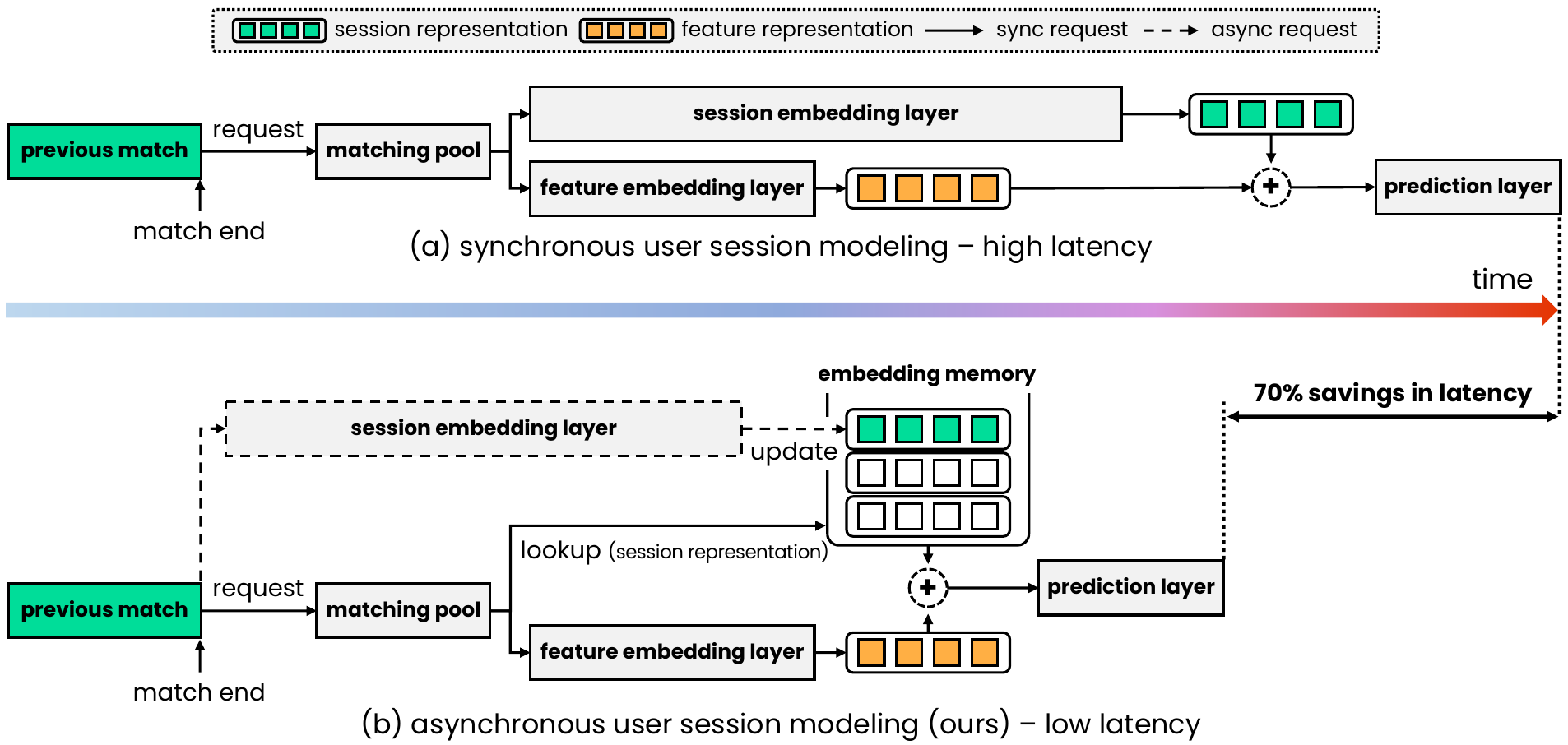}
\caption{System design consideration. (a) The overall latency of the session-based recommendation pipeline largely depends on the computational time of user session modeling (session embedding layer). (b) Our recommendation system, \textsc{Cupid}, reduces the latency of the recommendation pipeline by asynchronously conducting user session modeling in parallel with the pipeline.}
\label{fig:2_async_session_embedding}
\vspace{-6pt}
\end{figure*}

\section{Problem Formulation}
\label{sec:problem_formulation}

A real-time social discovery platform connects online users enabling immediate, one-on-one conversations. Let $\mathcal{U}$ represent the set of all users on such a platform. At any given time $t$, the matching pool $\mathcal{U}^{(t)} = \{u_1, u_2, \ldots, u_n\}$ consists of $n$ users available for matching. As illustrated in \autoref{fig:1_real_time_social_discovery}, the matching pool $\mathcal{U}^{(t)}$ is dynamic, constantly changing as users log in or out and conversations begin or end. Each user $u_i \in \mathcal{U}^{(t)}$ is characterized by a set of features $X_i$ (e.g., gender, country code, and other match-related statistics) and session information $S_i = [m_{i,1}, m_{i,2}, \ldots, m_{i,h}]$, which includes $h$ matching histories. Each matching history $m_{i,k} \in S_i$ consists of the chat counterpart $u_j$ and the chat duration $y_{ij}$ as follows:
\begin{equation}
    \textbf{Matching History:} \quad m_{i,k} = (u_i, u_j, y_{ij})\,.
\end{equation}

\noindent The goal of the session-based reciprocal recommendation system is to optimally pair suitable users from $\mathcal{U}^{(t)}$ by considering their features and matching histories to maximize overall user satisfaction. We define a recommendation model $f(\cdot)$, which estimates satisfaction scores $s_{ij}$ for all possible pairs of users $(u_i, u_j)$:
\begin{equation}
    \textbf{Satisfaction Score:} \quad s_{ij} = f(u_i, u_j)\,.
\end{equation}

\noindent For the satisfaction score, chat duration $y_{ij}$ is used as a proxy for satisfaction scores, based on the assumption that longer conversations correlate with higher user satisfaction. Therefore, the recommendation model's objective is revised to predict chat durations $\hat{y}_{ij}$ for each user pair:
\begin{equation}
    \textbf{Predicted Chat Duration:} \quad \hat{y}_{ij} = f(u_i, u_j)\,.
\end{equation}

\noindent These predictions are used to connect users through efficient matching algorithms designed according to the service's business logic. By predicting chat durations, the system can expedite connections that likely enhance user satisfaction, successfully addressing the challenges of dynamic user preferences in real-time social discovery platforms.

\section{Proposed Approach: Cupid}\label{sec:cupid}

In this section, we introduce \textsc{Cupid}, our session-based reciprocal recommendation system designed for real-world social discovery services with a focus on low-latency performance. We describe the implementation of Cupid and present a novel training method that efficiently captures mutual interests among users based on extensive matching histories.

\subsection{System Design Considerations}\label{subsec:4_1_system_design_considerations}

As highlighted earlier, delivering recommendations with minimal delay is crucial for real-time social discovery platforms. Any latency may lead to longer wait times for users, harming user experience and potentially causing them to leave the service. A key challenge is efficiently modeling short-term, dynamic user behaviors to capture real-time preferences and intents. \textsc{Cupid} addresses two primary considerations: (i) rapidly computing satisfaction scores for all potential user pairs in the matching pool to minimize latency, and (ii) overcoming the slower processing times associated with sequence modeling architectures, such as RNNs or transformers. To tackle these challenges, we have developed two core strategies for score computation and session modeling.

\vspace{4pt}
\noindent
\textbf{Linear Scaling Score Computation}\;\;
We compute the expected satisfaction score $\hat{y}_{ij}$ (i.e., chat duration) by applying a simple linear transformation to the dot product of user representations as follows:
\begin{equation}
\hat{y}_{ij} = f(u_i, u_j) = w(\mathbf{e}_i\cdot\mathbf{e}_j) + b,
\end{equation}
where $\mathbf{e}_i$ and $\mathbf{e}_j$ are the $d$-dimensional representation of users $u_i$ and $u_j$, respectively. This approach allows us to compute the matrix of predicted satisfaction scores $\hat{\mathbf{Y}}\in\mathbb{R}^{n\times n}$ for all user pairs efficiently using a single matrix multiplication by leveraging optimized BLAS\,\citep{blackford2002updated} libraries.

\begin{figure*}[t]
\centering
\includegraphics[width=0.9\textwidth]{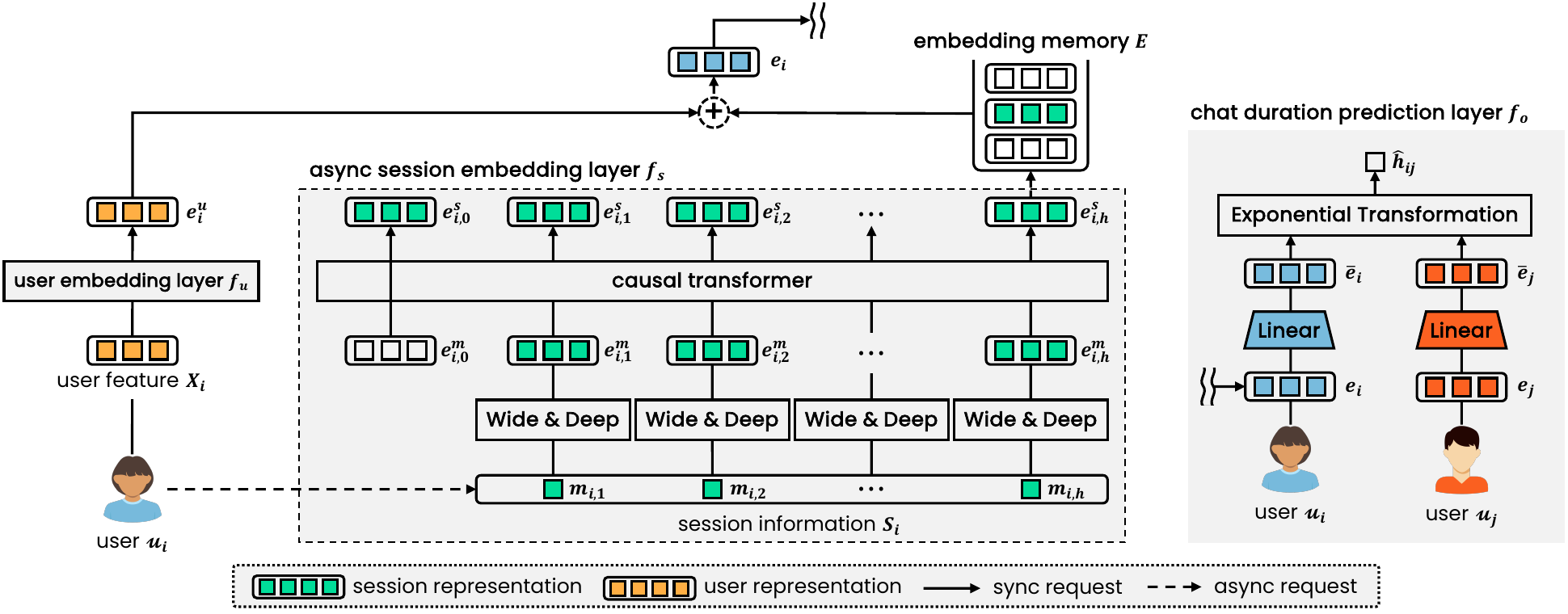}
\vspace{2pt}
\caption{An overview of \textsc{Cupid} architecture. The user $u_i$'s features $X_i$ and session information $S_i$ are modeled into the user feature representation $\mathbf{e}^u_i$ and the session representation $\mathbf{e}^s_i$ via the user feature embedding layer $f_u$ and the session embedding layer $f_s$, respectively. The session representation is asynchronously computed and stored in the embedding memory $E$.
}
\label{fig:3_overall_architecture}
\vspace{-4pt}
\end{figure*}

\vspace{4pt}
\noindent\textbf{Asynchronous Session Modeling}\;\; We decouple the computationally intensive user session modeling from the real-time matching pipeline by handling it asynchronously. This design significantly enhances the responsiveness of our recommendation system, enabling \textsc{Cupid} to deliver swift recommendations, which is essential for maintaining user engagement, as illustrated in \autoref{fig:2_async_session_embedding}. An overview of Cupid's architecture is provided in \autoref{fig:3_overall_architecture}. The performance of \textsc{Cupid} is measured by the \textit{Mean Squared Error (MSE)} as follows:

\begin{equation}
    \mathcal{L}_{\texttt{MSE}} = \frac{1}{|\mathcal{D}|}\sum_{m\in \mathcal{D}}(\hat{y}_{ij} - y_{ij})^2,
\label{eq:9_mse_loss}
\end{equation}
where $\mathcal{D}$ is the dataset match history of all users. Further details of \textsc{Cupid} are presented in the subsequent sections.

\subsection{Asynchronous Session Embedding Layer $f_s$}
\label{subsec:4_2_async_session_embedding_layer}
To ensure low-latency recommendations, \textsc{Cupid} models user behaviors in their sessions asynchronously rather than synchronously with matching requests. As illustrated in \autoref{fig:2_async_session_embedding}\textcolor{red}{(b)}, when a user $u_i$'s previous match ends, the session representation vector $\mathbf{e}^s_i$ is computed asynchronously using the session embedding layer $f_s$. More specifically, each matching history $m$ in user $u_i$'s session information $S_i$ is embedded into a representation $\mathbf{e}^m$ using Wide\& Deep model\,\citep{cheng2016wide}. This incorporates features $X_i$ from the user $u_j$, and the features $X_j$ from the chat counterpart user $u_j$, along with the chat duration $y_{ij}$. The user session representation $\mathbf{e}^s_i$ is then formed from these matching history representations [$\mathbf{e}^m_{i,1}, \mathbf{e}^m_{i,2},\cdots,\mathbf{e}^m_{i,h}$] employing a causal transformer, ensuring that each output $\mathbf{e}^s_{i,k}$ represents the user's state after the $k$-th match, influenced only by preceding matches. The final session representation $\mathbf{e}^s_i$ is stored in an embedding memory $E$, replacing any existing representation. When user $u_i$, requests a new match, the stored embedding $\mathbf{e}^s_i$ is retrieved to predict chat durations. Note that the session representation may not be updated before the session representation lookup occurs, as the computation might still be in progress when a new match is requested. In such cases, we refer to the session representation retrieved as a \textit{delayed} session representation.

This design provides significant advantages: it decouples the slower user session modeling from the synchronous matching pipeline, improving both recommendation speed and efficiency. However, asynchronously updating session representations may cause recent information to be displaced during inference, as new match data could arrive while the session representations are still being updated. Despite this, the system incurs only a few seconds of delay, so the impact on performance is negligible. Furthermore, by handling session information asynchronously, the overall throughput of session processing is enhanced through the batching of multiple inferences, which also reduces computational costs.

\subsection{Synchronous User Feature Embedding Layer $f_u$}
\label{subsec:4_3_sync_user_embedding_layer}
Along with session information, Cupid incorporates user features such as demographic details (e.g., gender, country) and other match statistics to capture general user preferences. We use Wide\&Deep\,\citep{cheng2016wide} as the user feature embedding layer $f_u$, which processes the user features $X_i$ to generate a representation $\mathbf{e}^u_i = f_u(X_i)$. This representation $\mathbf{e}^u_i$ is then used in the prediction layer to estimate chat duration of users.

\subsection{Chat Duration Prediction Layer $f_o$}
\label{subsec:4_4_chat_duration_prediction_layer}

The chat duration prediction layer $f_o$ aims to accurately predict the chat duration for a user pair ($u_i$, $u_j$) by combining their session and feature representations:
\begin{equation}
 \mathbf{e}_i = \mathbf{e}^s_i+\mathbf{e}^u_i, \:\:\:\: \mathbf{e}_j = \mathbf{e}^s_j+\mathbf{e}^u_j\,.
\label{eq:10_final_user_representation}
\end{equation}

While a simple method to predict chat duration may involve computing the dot product of these user representations ($\mathbf{e}_i \cdot \mathbf{e}_j$), this can lead to overestimating chat duration for users with similar profiles, resulting in sub-optimal recommendations when recommending similar users is not always ideal\,\citep{neve2020imrec,yang2022modeling}. To more accurately capture mutual interest while avoiding overestimation for similar users, we linearly project user representations into separate latent spaces:
\begin{equation}
    \Bar{\mathbf{e}}_i = \mathbf{W}_1\mathbf{e}_i+\mathbf{b}_1, \:\:\:\: \Bar{\mathbf{e}}_j = \mathbf{W}_2\mathbf{e}_j+\mathbf{b}_2,
\end{equation}
where $\mathbf{W}_1$ and $\mathbf{b}_1$ are the learnable weight matrix and bias for the projection of the representation of user $u_i$, and $\mathbf{W}_2$ and $\mathbf{b}_2$ are the corresponding weight matrix and bias for their chat counterpart $u_j$. Then, the predicted chat duration is estimated by the dot product of the mapped representations $\Bar{\mathbf{e}}_i$ and $\Bar{\mathbf{e}}_j$:
\begin{equation}
\hat{y}_{ij}=\Bar{\mathbf{e}}_i\cdot \Bar{\mathbf{e}}_j.
\label{eq:7_dot_product}
\end{equation}

\begin{figure}[t]
\centering
\includegraphics[width=0.95\columnwidth]{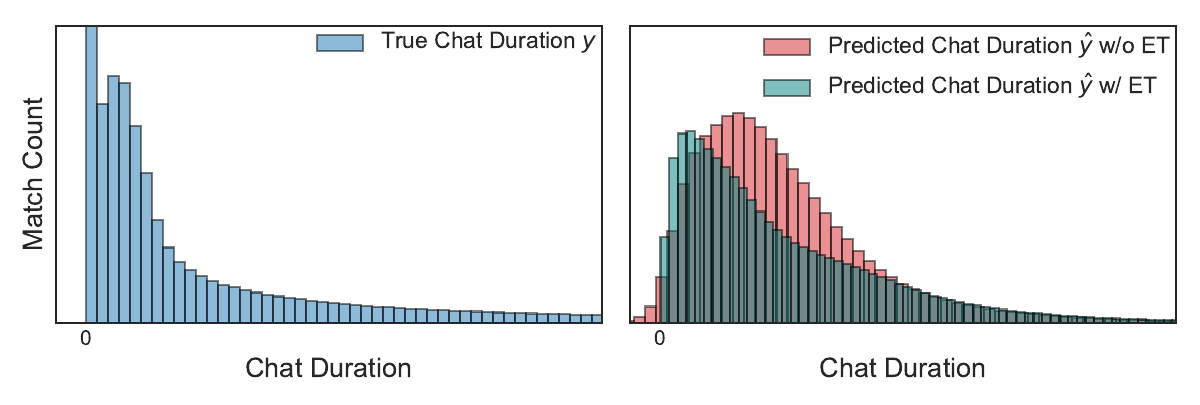}
\vspace{-4pt}
\caption{(Left): True chat duration distribution. (Right): Predicted chat duration \textit{with} and \textit{without} exponential transform.}
\label{fig:4_exponental_transform}
\vspace{-10pt}
\end{figure}

\noindent
\textbf{Exponential Transformation}\;\; As plotted in \autoref{fig:4_exponental_transform}, the actual chat durations follow a long-tailed distribution in real-world social discovery platforms (\textcolor{cyan!76!black}{\textbf{blue}} histogram) in practice. However, when trained with naive \textit{MSE} objective,
predictions based on the dot product tend to follow a normal distribution (\textcolor{red!70!white}{\textbf{red}} histogram), which deviates from the true distribution. To match predictions with the true distribution, we apply an exponential transformation to \autoref{eq:7_dot_product} as follows:
\begin{equation}
\hat{y}_{ij} = f_o(\mathbf{e}_i, \mathbf{e}_j) = \exp\left( w \left( \bar{\mathbf{e}}_i \cdot \bar{\mathbf{e}}_j \right ) + b \right), 
\label{eq:8_exponential_transformation}
\end{equation}
where $w$ and $b$ are learnable parameters.

As a result, the exponential transformation effectively adjusts the predicted durations to match the long-tailed distribution of actual chat durations (\textcolor{green!50!black}{\textbf{green}} histogram) with minimal computational overhead.

\subsection{Two-Phase Training}
\label{subsec:4_6_two_phase_learning}
\vspace{-4pt}

Training session-based recommendation systems in real-time contexts poses significant computational challenges due to the dynamic nature of user representations. Each user's preferences evolve after each interaction, requiring the system to frequently update their session representations to accurately reflect their current state. To make precise recommendations, the system must consider the updated session data for both users involved in each match. Traditionally, this involves processing and updating the session data for both the initiating user $u_i$ and their chat counterpart $u_j$ separately using complex models like transformers. For each user, the causal transformer processes their session history with a computational complexity of $O(|S|^2)$, where $|S|$ is the average session length per user. Modeling both users separately effectively doubles the computational cost, making it computationally intensive.

Moreover, accurately predicting matches requires jointly modeling how the sessions of $u_i$ and $u_j$ interact, which significantly increases computational overhead. This is because every interaction in $u_i$'s session might influence and be influenced by every interaction in $u_j$'s session, expanding the interaction space exponentially. In a naive approach, considering cross-attention between both users' sequences could lead to a theoretical complexity of $O(|S|^4)$. Such high computational demands make real-time processing prohibitive, especially in large-scale platforms with millions of users and high interaction rates like \textit{Azar}. The sequential dependencies inherent in causal transformers further exacerbate the issue, as each interaction's representation depends on all previous interactions, leading to extensive computations. To address this challenge and improve training efficiency, we propose a \textit{Two-Phase Training Strategy}, outlined in \textcolor{red}{Algorithm}\,\autoref{alg:1_two_phase_learning}, which significantly reduces computational overhead during training without substantially compromising the model's performance.

\begin{algorithm}[t]
\small
\begin{algorithmic}[1]
    \State {\bfseries Input:} feature embedding layer $f_u$, auxiliary feature embedding layer $\Tilde{f}_u$, session embedding layer $f_s$, and chat duration prediction layer $f_o$
    \State {\bfseries Output:} the trained layers $f_u$, $f_s$, and $f_o$
    \State {\textcolor{blue!60!gray}{\textbf{\textit{\# Phase\,1 Training (Embedding Layer)}}}}
    \Repeat 
    \For{$u_i$ $\in$ $\mathcal{U}$}
        \State compute $f_{s}(S_i)$ = [$\mathbf{e}^s_{i,0},\mathbf{e}^s_{i,1},\mathbf{e}^s_{i,2},\cdots,\mathbf{e}^s_{i,h}$]
        \For{$m_k = (u_i, u_{j}, y_{ij,k})$ $\in$ $S_i$}
        \State compute $\mathbf{e}^u_{i,k} = f_u(X_{i,k}),\:\:\Tilde{\mathbf{e}}^u_{j,k}=\Tilde{f}_u(X_{j,k})$
        \State compute $\mathcal{L}_{MSE}$ = $(f_o(\mathbf{e}^u_{i,k}+\mathbf{e}^s_{i,k-1},\Tilde{\mathbf{e}}^u_{j,k})-y_{ij,k})^2$
        \State update $f_u,\Tilde{f}_{u},f_s,f_o$ with $\mathcal{L}_{\texttt{MSE}}$
        \EndFor
    \EndFor
    \Until{CUPID converges}
    \State {\textcolor{blue!60!gray}{\textbf{\textit{\# Phase\,2 Training (Prediction Layer)}}}}
    \State freeze the feature embedding layer $f_u$ and session embedding layer $f_s$
    \State compute $\mathbf{e}^u_i,\mathbf{e}^s_i,\mathbf{e}^u_j,\mathbf{e}^s_j$ in advance
    \Repeat
    \For{$m=(u_i,u_j,y_{ij})\in \mathcal{D}$}

        \State compute $\mathcal{L}_{\texttt{MSE}} = (f_o(e^u_i+e^s_i, e^u_j+e^s_j) - y_{ij})^2$
        \State update $f_o$ with $\mathcal{L}_{\texttt{MSE}}$
    \EndFor
    \Until{CUPID converges}

\caption{Two-Phase Training Strategy}
\label{alg:1_two_phase_learning}
\end{algorithmic}
\end{algorithm}

\vspace{4pt}
\noindent
\textbf{Phase\,1: Training Embedding Layers}\;\;
The primary goal of this phase is to efficiently train the user feature embedding layer $f_u$ and the asynchronous session embedding layer $f_s$. We introduce an auxiliary user feature embedding layer $\tilde{f}_u$ to assist in training these layers excluding session information from the chat counterparts. This reduces the input to $(X_i; S_i, X_j)$, allowing us to leverage the causal transformer to generate session representations $\mathbf{e}^s_{i,k}$ with a single forward pass per user. This phase is aimed at minimizing the following objective:
\begin{equation}
\small
    \mathcal{L}_{\texttt{MSE}} = \frac{1}{|\mathcal{D}|}\sum_{u_i \in \mathcal{U}} \sum_{m_k \in S_i} \left( f_o\left( \mathbf{e}^u_{i,k} + \mathbf{e}^s_{i,k-1},\, \tilde{\mathbf{e}}^u_{j,k} \right) - y_{ij,k} \right)^2,
\end{equation}
where $\mathbf{e}^s_{i,k-1}$ is the session state of user $u_i$ after the $(k-1)$-th match for predicting the chat duration of the $k$-th match.

\vspace{4pt}
\noindent
\textbf{Phase\,2: Training the Chat Duration Prediction Layer}\;\;
In this phase, we enhance the chat duration prediction layer $f_o$ by fully incorporating session information from both users in each match $(X_i; S_i, X_j; S_j)$. Thereby, the objective becomes:
\begin{equation}
\small
    \mathcal{L}_{\texttt{MSE}} = \frac{1}{|\mathcal{D}|}\sum_{u_i \in \mathcal{U}} \sum_{m_k \in S_i} \left( f_o\left( \mathbf{e}^u_{i,k} + \mathbf{e}^s_{i,k-1},\, \mathbf{e}^u_{j,k} + \mathbf{e}^s_{j,k-1} \right) - y_{ij,k} \right)^2.
\end{equation}
In this final phase, we discontinue using the auxiliary user feature embedding layer $\tilde{f}_u$ from the first phase and freeze the embedding layers ($f_u$, $f_s$) to improve processing efficiency. By computing the user feature representations in advance, subsequent calculations can be optimized.

\vspace{4pt}
\noindent
\textbf{Computational Complexity Analysis}\;\; We analyze how our two-phase training strategy enhances training efficiency, as detailed in \autoref{tab:3_num_transformer_inference}. Let $N$ denote the total number of training epochs in standard learning, with $N_1$ and $N_2$ for the first and second phases, respectively. $|\mathcal{D}|$ denotes the number of matching histories in the dataset, and $\overline{|S|}$ indicates the average session length per user. In standard training, modeling session representations for both users requires $2N|\mathcal{D}|$ inferences since both $u_i$ and $u_j$ need to be processed for each match history across all epochs. In contrast, the first phase of our method requires only $N_1 |\mathcal{D}| / \overline{|S|}$ inferences, as we compute session representations for only one user per inference and leverage the average session length to reduce computations. During the second phase, by pre-extracting user representations and freezing the embedding layers $f_s$ and $f_u$, the total number of inferences needed is just $2|\mathcal{D}| / \overline{|S|}$, regardless of $N_2$. This method reduces the inferences required by the causal transformer in $f_s$ to:
\[
\frac{2N \overline{|S|}}{N_1 + 2}
\]
compared to $2N|\mathcal{D}|$ in conventional methods. Assuming $N = N_1 = 10$ and $\overline{|S|} = 128$, our method achieves a \textbf{213x reduction} in transformer inferences. Considering that transformer inference constitutes the majority of training latency, this substantial reduction greatly facilitates the efficient training of \textsc{Cupid}, even with large datasets.
\begingroup
\setlength{\tabcolsep}{8pt} 
\renewcommand{\arraystretch}{1.0}
\begin{table}[t!]
\small
\caption{Comparison of the number of causal transformer inferences with and without our two-phase training.}
\vspace{-1pt}
\begin{tabular}{c|c|c|c}
\toprule
\multicolumn{2}{c|}{\textbf{Phase}} & \multicolumn{2}{c}{\textbf{Time Complexity}} \\
\midrule\midrule
\multicolumn{2}{c|}{w/o Two-phase} & \multicolumn{2}{c}{$2N|\mathcal{D}|$} \\
\hline
\multirow{2}{*}{w/ Two-phase} & Phase-1 & 
$N_1|\mathcal{D}|/\overline{|S|}$ & \multirow{2}{*}{$(N_1+2)|\mathcal{D}|/\overline{|S|}$}\\
\cline{2-3} 
& Phase-2 & $2|\mathcal{D}|/\overline{|S|}$ &  \\
\midrule
\multicolumn{2}{c|}{Reduction Factor} & \multicolumn{2}{c}{$2N \overline{|S|}/(N_1+2)$} \\
\bottomrule
\end{tabular}
\label{tab:3_num_transformer_inference}
\vspace{-8pt}
\end{table}
\endgroup
\section{Experiments}
\label{sec:experiments}

\subsection{Experimental Setups}

\vspace{4pt}
\noindent\textbf{Data Setups}\;\; The performance of \textsc{Cupid} is evaluated in both offline and online environments. In the \textit{offline} evaluation, its performance is tested in a controlled setting. A large-scale matching history from \textit{Azar} is used, consisting of a billion-scale dataset from millions of user sessions generated over a month. Data from the last two days is used for validation and testing, while the remaining data is used as the training set. In the \textit{online} evaluation, \textsc{Cupid}'s effectiveness is validated in real-world conditions to ensure that the gains observed are consistent in a live service environment.

\vspace{4pt}
\noindent\textbf{Evaluation Setups}\;\; We adopt two baseline models based on Wide\&Deep\,\citep{cheng2016wide}, which were previously used in \textit{Azar} before adopting session-based recommendations as follows:
\begin{itemize}[leftmargin=2pt, labelwidth=0pt, itemindent=2pt, align=left]
    \item \textbf{Wide\&Deep}: 
    A widely adopted recommendation method that captures higher-order interactions among input features using neural networks. It employs user representations $e_i = e_i^u$ and $e_j = e_j^u$ are employed in \autoref{eq:10_final_user_representation} without session representations $e_i^s$ and $e_j^s$. Consequently, it relies solely on static user features and does not include any real-time information from user sessions.
    \item \textbf{Wide\&Deep-S}: A variant of Wide\&Deep that incorporates real-time features generated during user sessions. It captures user behaviors while maintaining low latency by leveraging aggregated features from recent match histories, such as average chat duration, along with existing user features. This baseline serves to demonstrate the effectiveness of our sequential approach for modeling user sessions.
\end{itemize}

For performance evaluation, we use \textit{MSE} and \textit{Area Under the Receiver Operating Characteristic (AUROC)}. \textit{MSE} measures the average squared difference between actual and predicted chat durations by applying log-scaled chat durations (\texttt{ms}) to minimize the impact of noise in shorter intervals. The same log-scaling is also used during the training of our recommendation models. In contrast, \textit{AUROC} assesses the model's ability to distinguish between potential matches that result in quality interactions and those that do not. A quality match is defined as one where the chat duration exceeds a specific threshold. As latency is another critical factor for real-time reciprocal recommendation, we also evaluate the latency improvement achieved by adopting \textsc{Cupid} in the real-world deployment of \textit{Azar}.
\begin{figure}[t]
\centering
  \centering
  \resizebox{\columnwidth}{!}{\includegraphics[width=0.64\textwidth]{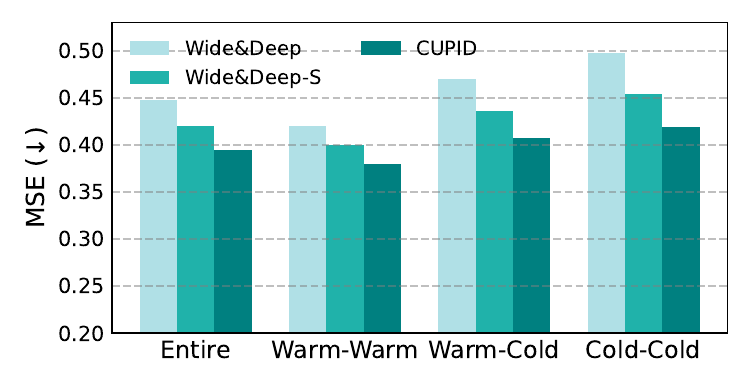}
  }
  \vspace{-6pt}
  \centering
  \resizebox{\columnwidth}{!}{\includegraphics[width=0.64\textwidth]{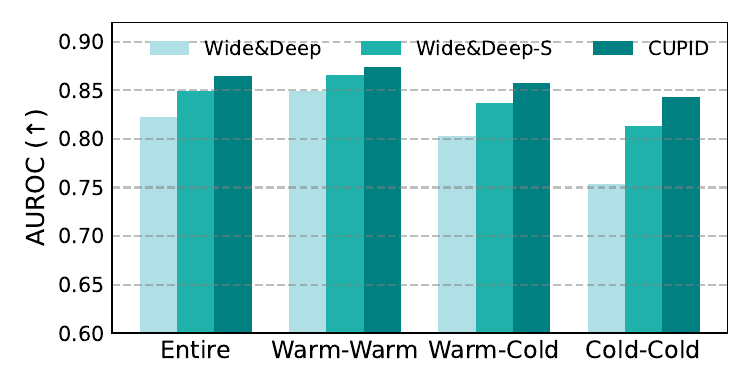}}
  \vspace{-11pt}
\caption{\textit{Offline} evaluation results on four types of matches.}
\label{fig:5_main_results}
\end{figure}
\begingroup
\setlength{\tabcolsep}{4pt} 
\renewcommand{\arraystretch}{1.12}
\begin{table}[t]
    \begin{center}
    \small
    \vspace{-3pt}
    \caption{
    \textit{Online} evaluation results across all user segments in \textit{Azar} production. The relative performance change of \textsc{Cupid} compared to the baseline Wide\&Deep is reported.
    }
    \begin{tabular}{c|ccc}
    \toprule
    \textbf{User Segment}  & \textbf{\makecell{Average \\Chat Duration}} & \textbf{\makecell{Long Match\\Ratio}} & \textbf{\makecell{Short Match\\ Ratio}} \\
    \midrule
    All users        & +6.8\%                & +12.6\%          & -2.4\%            \\ 
    Warm-start users        & +6.8\%                & +12.9\%          & -2.3\%            \\ 
    Cold-start users        & +5.9\%                & +9.7\%          & -4.1\%            \\ 
        \bottomrule
    \end{tabular}%
    \label{tab:4_live_test_result}
    \end{center}
    \vspace{-16pt}
\end{table}
\endgroup

\begingroup
\setlength{\tabcolsep}{6pt} 
\renewcommand{\arraystretch}{1.05}
\begin{table*}[t]
    \begin{center}
    \small 
    \caption{The simulation of the deployment environment where user session representations may not up-to-date. We observe the performance changes by using delayed user session representations stored at time $(t-t')$, with varying delay time $t'$.}
    \begin{tabular}{c|c|cc|cc|cc|cc}
        \toprule
        \multirow{2}{*}{\textbf{Method}}&\multirow{2}{*}{\textbf{$t'$}} &\multicolumn{2}{c|}{\textbf{Entire Match}}& \multicolumn{2}{c|}{\textbf{Warm-Warm Match}}&\multicolumn{2}{c|}{\textbf{Warm-Cold Match}}& \multicolumn{2}{c}{\textbf{Cold-Cold Match}}  \\\cmidrule{3-10}
        & & MSE ($\downarrow$) &  AUROC ($\uparrow$)&  MSE ($\downarrow$)&  AUROC ($\uparrow$)&  MSE ($\downarrow$)&  AUROC ($\uparrow$)&  MSE ($\downarrow$)&  AUROC ($\uparrow$)\\
        \midrule\midrule
         \multirow{5}{*}{\textsc{Cupid}} & - &\textbf{0.3968}&\textbf{0.8635}&\textbf{0.3815}&\textbf{0.8735}& \textbf{0.4094}&\textbf{0.8564}&\textbf{0.4214}&\textbf{0.8409} \\
          & 2000\texttt{ms}&0.3989&0.8616&0.3834&0.8719&0.4117&0.8541&0.4239&0.8389 \\
          & 4000\texttt{ms}& 0.3990 & 0.8616 & 0.3834&0.8719&0.4118&0.8540&0.4239& 0.8389\\
          & 8000\texttt{ms}&0.3993&0.8614&0.3837&0.8717&0.4121&0.8538&0.4242&0.8386 \\
          & 16000\texttt{ms}&0.4004&0.8605&0.3848&0.8710&0.4132&0.8528&0.4254&0.8375 \\
          \midrule
        Wide\&Deep-S & -&0.4197&0.8497&0.3996&0.8655&0.4359&0.8375&0.4539&0.8136 \\
        \bottomrule
    \end{tabular}%
    \label{tab:2_delayed_session_representation}
    \end{center}
    \vspace{-2pt}
\end{table*}
\endgroup


\begingroup
\setlength{\tabcolsep}{8.0pt} 
\renewcommand{\arraystretch}{0.95}
\begin{table*}[t!]
    \begin{center}
    \small 
    \caption{Ablation test results. 
    SP and ET denote the second phase in our two-phase learning and the exponential transform in \autoref{eq:8_exponential_transformation}, respectively. 
    The performance changes are observed by removing each component.
    }
    \begin{tabular}{ccc|cc|cc|cc|cc}
        \toprule
        \multicolumn{3}{c|}{\textbf{Components}} &\multicolumn{2}{c|}{\textbf{Entire Match}}& \multicolumn{2}{c|}{\textbf{Warm-Warm Match}}&\multicolumn{2}{c|}{\textbf{Warm-Cold Match}}& \multicolumn{2}{c}{\textbf{Cold-Cold Match}}  \\\midrule
        $\mathbf{e}^s$& SP& ET& MSE ($\downarrow$) &  AUROC ($\uparrow$)&  MSE ($\downarrow$)&  AUROC ($\uparrow$)&  MSE ($\downarrow$)&  AUROC ($\uparrow$)&  MSE ($\downarrow$)&  AUROC ($\uparrow$)\\
        \midrule\midrule
         \XSolidBrush&\CheckmarkBold&  \CheckmarkBold&0.4197   &0.8497  &0.3996 &0.8655& 0.4359&0.8375&0.4539&0.8136 \\
       \CheckmarkBold&\XSolidBrush&\CheckmarkBold&0.4271&0.8464&0.4059&0.8649&0.4436&0.8329&0.4648&0.7995 \\
         \CheckmarkBold&\CheckmarkBold&\XSolidBrush& 0.3996 & 0.8615 & 0.3845&0.8712&0.4121&0.8545&0.4239& 0.8394\\          \midrule
         \CheckmarkBold&  \CheckmarkBold& \CheckmarkBold &\textbf{0.3948}&\textbf{0.8648}&\textbf{0.3797}&\textbf{0.8745}& \textbf{0.4072}&\textbf{0.8577}&\textbf{0.4190}&\textbf{0.8431} \\
        \bottomrule
    \end{tabular}%
    \label{tab:1_ablation_test}
    \end{center}
    \vspace{-12pt}
\end{table*}
\endgroup

\subsection{Offline Performance Evaluation}

In \autoref{fig:5_main_results}, the overall performances across three match types are presented. The match types include \textit{Entire} Match, which encompasses all categories; \textit{Warm-Warm}, for matches between warm-start users; \textit{Warm-Cold}, for matches between warm-start and cold-start users; and \textit{Cold-Cold}, for matches exclusively between cold-start users. Here, cold-start users have no previous matching history in the training dataset. The distribution is 58.1\% for \textit{Warm-Warm}, 35.5\% for \textit{Warm-Cold}, and 6.3\% for \textit{Cold-Cold}. \textsc{Cupid} consistently outperforms baseline methods across all categories and metrics. 

\subsection{Online Production Performance}
\begingroup
\setlength{\tabcolsep}{11.6pt} 
\renewcommand{\arraystretch}{1.05}
\begin{table}[t]
    \begin{center}
    \small 
    \caption{
    Response latency of \textsc{CUPID} in online environments under the {\textit{Azar}} service scenario.
    }
    \begin{tabular}{c|cc}
    \toprule
    \multirow{2}{*}{\textbf{Components}}  & \textbf{90-th} & \textbf{99-th} \\
     & \textbf{percentile} & \textbf{percentile} \\
    \midrule\midrule
    User representation $\mathbf{e}^u$ & 9ms & 17ms \\ 
    
    Session representation $\mathbf{e}^s$ & 236ms & 290ms \\ 
    \cline{1-3}
    Synchronous implementation & 236ms & 290ms \\
    \midrule
    \textsc{CUPID}:Asynchronous & 48ms & 70ms\\
     implementation \textbf{(Ours.)} & \textcolor{green!50!black}{\textbf{(-79.7\%)}} & \textcolor{green!50!black}{\textbf{(-75.9\%)}} \\
    \bottomrule
    \end{tabular}
    \label{tab:5_online_latency}
    \end{center}
\vspace{-15pt}
\end{table}
\endgroup

While \textsc{Cupid} shows a significant improvement in predicting satisfaction scores in offline experiments, it may not always lead to increased user engagement online. To evaluate its real-world impact, we test \textsc{Cupid} in the production environment of \textit{Azar}, comparing it with the baselines. We conduct a Switchback\,\citep{robins1986switchback} test instead of an A/B test due to the shared matching pool, which makes it difficult to independently separate A/B groups. The results in \autoref{tab:4_live_test_result} show improvements in metrics such as average chat duration and the ratio of long to short matches, defined by a preset threshold. For all user segments, \textsc{Cupid} consistently increases the average chat duration and improves match quality. This demonstrates that \textsc{Cupid} not only accurately predicts satisfaction scores but also enhances user experience in a live setting. Meanwhile, \autoref{tab:5_online_latency} shows the latency improvement achieved by \textsc{Cupid}, emphasizing its primary goal of delivering low-latency recommendations through asynchronous session modeling. In the real-world deployment of \textit{Azar}, \textsc{Cupid} reduces latencies at the 90th and 99th percentiles by up to 79.7\% compared to synchronous computation of session representations in the matching pipeline. This significant reduction ensures stable latency, which is essential for real-time services.

\subsection{Effect of Delayed Session Representation}\label{subsec:5_5_effect_of_delayed_session_representation}
In real deployment, the session representation $\mathbf{e}^s$ might miss the latest matching histories if a user requests a new match before the update is complete. The system then uses a delayed representation, which lacks data from the most recent matches. To study the impact of this delay, we simulate an environment where the representation update is delayed for $t'$ milliseconds and predict chat durations for users in the matching pool $\mathcal{U}^{(t)}$ using this delayed data. The results are summarized in \autoref{tab:2_delayed_session_representation}. Two main observations emerge. First, prediction performance slightly decreases as delay time increases, which is expected since the system design decouples session modeling from the synchronous matching pipeline to avoid latency issues. This compromise is acceptable, as it prevents session modeling from becoming a bottleneck. Second, even with this delay, the models still outperform the Wide\&Deep-S baseline by a significant margin in all cases while maintaining similar latency. This shows that the approach, with its decoupled session modeling, achieves an optimal balance between latency and prediction performance.

\subsection{Ablation Study}
An ablation test is conducted to evaluate the impact of individual components on \textsc{Cupid}'s performance, focusing on session representation $\mathbf{e}^s$, the Exponential Transformation (ET), and the second phase of the two-phase training strategy. The results, shown in \autoref{tab:1_ablation_test}, indicate a performance drop when any component is removed. Excluding the session representation results in a significant decline, especially for cold-start users, underscoring its role in capturing mutual interests. Skipping the second-phase training also negatively impacts performance, highlighting its importance in using session data from both users to improve chat duration predictions. Additionally, omitting the exponential transformation leads to poorer performance, underscoring its value in aligning predicted chat durations with the actual distribution and stabilizing model training.
\section{Related Work}
\label{sec:related_work}

\noindent\textbf{Reciprocal Recommendation}\;\; Reciprocal recommendation systems differ from conventional in the sense of they aim to enhance mutual satisfaction through user-to-user recommendations, instead of focusing on item-to-user recommendations~\citep{palomares2021reciprocal,abdollahpouri2020multistakeholder,palomares2020reciprocal}. These systems have been widely studied, especially in contexts like online dating~\citep{neve2019latent,tomita2022matching, alanazi2013people,tu2014online}, and job search platforms~\citep{jiang2020learning,lu2013recommender,yang2022modeling,yan2019interview}. Our work shifts the focus to real-time reciprocal recommendations, where candidates appear and disappear dynamically. This is the first comprehensive study to investigate these complex dynamics in real-time.

\vspace{4pt} \noindent\textbf{Session-Based Recommendation}\;\; Session-based recommendation systems predict the next item by capturing dynamic user behaviors and intents within a session. Various models, such as Markov Chains~\citep{rendle2010factorizing,he2016fusing}, recurrent neural networks~\citep{hidash2016session,jannach2017recurrent,liu2020long}, graph neural networks~\citep{wu2019session,guo2022learning,zhang2023efficiently,pang2022heterogeneous}, transformers~\citep{vaswani2017attention,sun2019bert4rec,de2021transformers4rec,xia2023efficient,zhou2020s3}, and other attention mechanisms~\citep{li2017neural,liu2018stamp,zhou2022filter} have been utilized for this purpose. Our study extends session-based recommendations into the underexplored area of reciprocal recommendation tasks. While \cite{zheng2023reciprocal} examines sequential recommendations in a two-sided market, it does not address the low-latency requirements essential for real-time one-on-one social discovery platforms. In contrast, our work specifically focuses on meeting these extreme low-latency constraints, facilitating rapid and efficient user matching in reciprocal session-based recommendation systems.
\section{Conclusion}
\label{sec:conclusion}
To the best of our knowledge, this is the first study to develop a session-based reciprocal recommendation system optimized for real-time social discovery platforms. Our approach tackles stringent latency requirements by using asynchronous session modeling, which significantly reduces the time required for processing. Additionally, we introduce an efficient two-phase training method that simplifies the complexities of combining session-based and reciprocal recommendations. Our system, validated on a large-scale offline dataset and in a real-world environment, increases average chat duration by 6.8\% for warm-start users and 5.9\% for cold-start users. Moreover, it achieves over a 76\% reduction in latency compared to purely synchronous methods. This research opens a new direction for session-based real-time reciprocal recommendations.

\vspace{8pt}
\noindent\textbf{Ethical Statement}\;\;
By introducing \textsc{Cupid}, we aim to enhance user engagement and satisfaction through efficient, personalized matchmaking in social discovery. Using asynchronous session modeling and a two-phase training strategy, \textsc{Cupid} addresses low latency and dynamic user preferences. However, deploying such a system involves ethical considerations, including user privacy, data security, and potential algorithmic biases. To address these, we ensure strict adherence to data protection laws, implement robust security measures, and commit to developing fairness-aware algorithms with regular audits to prevent unintended discrimination.



\bibliography{main}
\bibliographystyle{IEEEtran}

\end{document}